\documentclass[twocolumn,floatfix,prb,showpacs,superscriptaddress]{revtex4-1}
\usepackage{helvet}
\usepackage{eurosym}
\usepackage{color}
\usepackage[latin1]{inputenc}     
\usepackage{subfigure}
\usepackage{amssymb}
\usepackage{amsmath}
\usepackage{graphicx}
\usepackage{bbold}
\usepackage{natbib}

\usepackage{hyperref}
\hypersetup{breaklinks}

\newcommand{\abs}[1]{\left| #1 \right|} 
\newcommand{\avg}[1]{\left< #1 \right>} 
\newcommand{\pd}[2]{\frac{\partial #1}{\partial #2}}

\newcommand{\ket}[1]{\left| #1 \right>} 
\newcommand{\ketbra}[2]{\left| #1 \vphantom{#2} \right>
 \left< #2 \vphantom{#1} \right|} 

\begin{document}
\title{Counting statistics of the Dicke superradiance phase transition}
\author{Wassilij Kopylov}
\author{Clive Emary}
\author{Tobias Brandes}
\affiliation{Institut f\"ur Theoretische Physik,
  Technische Universit\"at Berlin,
  D-10623 Berlin,
  Germany}
\date{\today}

\begin{abstract}
We consider a driven single mode Dicke-Hamiltonian coupled to a dissipative zero-temperature bath. We derive the cumulant generating function for emitted photons of this quantum-critical system by using a $P$-representation of the master equation in the thermodynamic limit.
This cumulant generating function is shown to consist of two parts: a macroscopic  component, which is Poissonian in nature with characteristic rate proportional to the order parameter of the system; and a part describing fluctuations which is non-trivial in form and divergent around the quantum phase transition.
\end{abstract}
\maketitle


\section{Introduction}
The Dicke model\cite{Dicke-Dicke_Modell} describes the interaction of $N$ two-level systems with bosonic field modes. In the thermodynamic limit, and for a single bosonic mode, it undergoes a quantum phase transition (QPT) from a normal to a superradiant phase when the atom-field coupling  strength exceeds a critical value\cite{Wang-phasetransition_dicke_modell,Hepp1973360,Chaos_and_quantum_phase_Dicke_modell_Clive_Brandes,PhysRevLett.90.044101}. The model has been experimentally realised by an  Bose-Einstein-condensate trapped in an optical cavity\cite{Baumann-Dicke_qpt,Nagy-dicke_and_bose_einstein}.  Dissipation due to photon emission has furthermore been modeled, e.g. by Heisenberg-Langevin equations \cite{Nagy_critical_exponent_open_dicke_model,Keeling_Collective_dynamics_of_BEC,oztop-excitations_of_otpically_driven_atomic_condensata_theory_photodetection}, the Keldysh approach \cite{torre_keldysh_phase_transition_dicke} or by use of Hartree-Fock-Bogoliubov theory \cite{Nagy-Domokos-Finite_size_scaling_qpt_open_dicke_model}.

In this paper, we study the full photon counting statistics of the driven dissipative single-mode Dicke model by including a counting field $\chi$ in the master equation, which we analytically solve in the $P$-representation in the thermodynamic limit $N \to \infty$. We obtain analytic expressions for time-dependent cumulants of the photon counting statistics, as well as the asymptotic cumulant generating function (CGF). The CGF is shown to consist of two parts: a macroscopic  component, which is Poissonian in nature and has a characteristic rate proportional to the mean occupation of the cavity mode (order parameter of the system); on top of this comes a contribution describing fluctuations about the mean behaviour, which has a non-trivial form and is divergent around the quantum phase transition.
In addition, we identify the three phases (normal, superradiant, intermittent) and the corresponding critical coupling parameters $\lambda_1,\lambda_2,\lambda_3$  that characterize the dissipative phase transition.

The structure of this paper is as follows: In Sec.~\ref{ch:Model} we introduce the model and obtain the three critical coupling parameters, in Sec.~\ref{ch:mseq_p_represent_counting} we solve the master equation with a counting field $\chi$, in Sec.~\ref{ch:results} we give an analytical expression for the CGF and in Sec.~\ref{ch:discussion} we discuss our results and connect them to other works.

\section{Model}
\label{ch:Model}
The Dicke Hamiltonian  ($\hbar =1$)
\begin{equation}\label{eq1:h0_origin}
 H = \omega_0 J_z + \omega a_1^\dag a_1 + \frac{\lambda}{\sqrt{2j}}(a_1^\dag + a_1)(J_+ + J_-)
\end{equation}
describes our isolated system. Here, $J_z,J_\pm$ are the collective atomic angular momentum operators that describe an ensemble of $N$ two level atoms with a level splitting $\omega_0$;
$j$ is the length of pseudo-spin with value $j = N/2$ for bosonic realisation \cite{Alcade_Bucher_Clive-Thermal_phase_transitions_for_dicke_type_models}, $a_1$ and $a_1^{\dag}$ are the ladder operators for the optical mode with an energy $\omega$, and $\lambda$ is the coupling strength between the optical mode and the atoms.
As in Refs.~\cite{Baumann-Dicke_qpt} we interpret Eq. \eqref{eq1:h0_origin} as an effective model in a frame rotating at the frequency of an external driving laser, $\Omega$.

We describe dissipation in this system with the master equation in Lindblad-form \cite{torre_keldysh_phase_transition_dicke}
\begin{align}\label{eq1:msgl_nagy_form}
 \frac{d}{dt} \rho(t) &= -i[H,\rho(t)] \notag \\
        &\quad - \frac{\Gamma}{2} \Biggl{[}
       a_1^\dag  a_1 \rho(t)  -  2 a_1 \rho(t) a_1^\dag + \rho(t)  a_1^\dag a_1
    \Biggr{]}  \,,
\end{align}
where $\rho$ is the density matrix of the atom-cavity system and $\Gamma$ is the rate of photon loss from the cavity.
\footnote{
  With bath Hamiltonian $H_B=\sum_k \omega_k b_k^\dag b_k$, the master equation \eqref{eq1:msgl_nagy_form}
  is obtained from a cavity-bath coupling Hamiltonian
  $
     H_{SB}(t) = \sum_k g_k (a_1^\dag b_k e^{i\Omega t} + b_k^\dag e^{-i \Omega t} a_1)
  $,
  in a frame rotating with driving frequency $\Omega$.
  This also means, that $\lambda$ is a function of $\Omega$.
} \\
To make analytical progress, we use the Holstein-Primakoff-transformation \cite{holstein-primakoff}, which allows one to describe $N$-two-level-atoms by a single bosonic mode. This is done by the transformation
\begin{eqnarray}\label{eq1:Holstein_primakoff_trafo}
    J_+ &=& a_2^\dag \sqrt{2j - a_2^\dag a_2}; \quad
    J_- = \sqrt{2j - a_2^\dag a_2}\,a_2;
    \nonumber \\
    J_z &=& (a_2^\dag a_2 - j)
    ,
\end{eqnarray}
where $a_2$ and $a_2^{\dag}$ are the ladder operators of the introduced atomic mode. In this representation, the Dicke-Hamiltonian reads
\begin{eqnarray}\label{eq2:Gesamthamiltonian_allgemein}
 H &=& \omega_0 (a_2^\dag a_2 - j) + \omega a_1^\dag a_1  \notag\\
        &&
        \!\!\!\!\!\!
        + \lambda (a_1^\dag + a_1)\left(a_2^\dag \sqrt{1 - \frac{a_2^\dag a_2}{2j}} + \sqrt{1 - \frac{a_2^\dag a_2}{2j} }a_2\right)
        \!\!.
\end{eqnarray}

\subsection{Thermodynamic Limit $N \to \infty$}
The isolated Dicke-Hamiltonian $H$ in the thermodynamic limit  $N \to \infty$ is known to have two phases: normal and superradiant \cite{Chaos_and_quantum_phase_Dicke_modell_Clive_Brandes}.

In the normal phase and for $N \to \infty$ only $H$ becomes effectively\cite{Chaos_and_quantum_phase_Dicke_modell_Clive_Brandes},
\begin{equation}\label{eq1:H0_NR}
  H_{N} = \omega_0 a_2^\dag a_2 + \omega a_1^\dag a_1 + \lambda(a_2^\dag + a_2)(a_1^\dag + a_1),
\end{equation}
which has the form of two simply-coupled harmonic oscillators.  The master equation for this phase may be obtained by simply replacing $H$ in Eq. \eqref{eq1:msgl_nagy_form} with this form.\\
In contrast, the superradiant phase is characterised by the macroscopic occupation of the $a_1$ and $a_2$ modes. In order to describe this phase, we insert a mean-field-ansatz
\begin{equation} \label{eq1:Mean_field_ansatz}
a_1 = c  + \sqrt{\alpha}, \, a_1^\dag = c^\dag + \sqrt{\alpha^*},\,
a_2 = d - \sqrt{\beta}, \,a_2^\dag = d^\dag - \sqrt{\beta^*}\,,
\end{equation}
into both $H$ and the dissipator term of the master equation, Eq. \eqref{eq1:msgl_nagy_form}.
Let us start with $H$, where we insert Eq. \eqref{eq1:Mean_field_ansatz} into Eq. \eqref{eq2:Gesamthamiltonian_allgemein}.  We then expand the square roots and neglect terms with powers of $N$ in the denominator \cite{Chaos_and_quantum_phase_Dicke_modell_Clive_Brandes}.  Assuming $\beta \in \mathbb{R}$ we obtain in the thermodynamic limit
\begin{eqnarray}\label{eq1:H0_SR_mit H01_und_H02}
  H_{S} = H_{S}^{(1)} + H_{S}^{(2)},
\end{eqnarray}
with
\begin{eqnarray}
  \label{eq1:H0_SR_mit H01}
  H_{S}^{(1)} &=&
  -2 \lambda \sqrt{\frac{k}{2j}} (c^\dag+ c) \sqrt{\beta} + \omega (\sqrt{\alpha} c^\dag + \sqrt{\alpha^*}c ) \notag\\
  &&
  + \lambda \sqrt{\frac{k}{2j}} (d + d^\dag) (\sqrt{\alpha} + \sqrt{\alpha^*})(1 - \frac{\beta}{k}) \notag\\
  &&
  - \omega_0 \sqrt{\beta} (d^\dag + d);
  \\
  \label{eq1:H0_SR_mit H02}
  H_{S}^{(2)} &=&
  \omega c^\dag c + \Omega_0 d^\dag d + \Lambda (c^\dag + c)(d^\dag + d)
  \notag\\
  &&
  + M (d^\dag + d)^2,
\end{eqnarray}
with
\begin{eqnarray}
  k &=& 2j - \beta;
  \nonumber \\
  \Omega_0 &=& \omega_0 + \lambda \sqrt{\frac{k}{2j}}
    (\sqrt{\alpha} + \sqrt{\alpha^*})\frac{\beta}{k}\,;
  \notag\\
  \Lambda &=& \lambda \sqrt{\frac{k}{2j}} \cdot \left(1 - \frac{\beta}{k}\right)\,;
  \notag\\
  \label{eq1:H2-parameter}
  M &=&  \lambda \sqrt{\frac{k}{2j}}(\sqrt{\alpha} + \sqrt{\alpha^*}) \frac{\sqrt{\beta}}{2} \frac{2k + \beta}{2k^2} \,.
\end{eqnarray}

Making the displacements in the dissipator, the complete master equation then reads
\begin{align}\label{eq1:msgl_nagy_form_mean_field_trafo}
    \frac{d}{dt} \rho(t) &= - i\left[H_{S}^{(1)} - i \frac{\Gamma}{2} \sqrt{\alpha} c^\dag + i \frac{\Gamma}{2} \sqrt{\alpha^*} c ,\rho(t) \right] \notag\\
& - i \left[H_{S}^{(2)},\rho(t)\right]
    - \frac{\Gamma}{2} \left(c^\dag c \rho(t) + \rho(t) c^\dag c - 2 c \rho(t) c^\dag\right)\,.
\end{align}
Now, we first determine the macroscopic occupation $\abs{\alpha}, \abs{\beta}$ in the superradiant phase using the master equation. In Eq. \eqref{eq1:msgl_nagy_form_mean_field_trafo} we have already separated the terms that are linear in the new operators $c, c^\dag$. These terms are proportional to the square root of the particle number $N$, so they diverge in the thermodynamic limit. We determine the macroscopic occupation, such that all parts $\sim \sqrt{N}$ of the master equation vanish. That means the first commutator has to be zero which is fulfilled if
\begin{equation}
H_{S}^{(1)} - i \frac{\Gamma}{2} \sqrt{\alpha} c^\dag + i \frac{\Gamma}{2} \sqrt{\alpha^*} c =0 \,.
\end{equation}
Next, we insert the $H_{S}^{(1)}$ from Eq. \eqref{eq1:H0_SR_mit H01}, factor out the operators $c, c^\dag, d, d^\dag$ and get an expression of the form $c (..) + c^\dag(..) + d(..) + d^\dag(..)=0$. To achieve the identity we set all braces to zero and arrive at four equations. The last two of them are equal, because we have assumed $\beta \in \mathbb{R}$. These equations can be solved (see  Appendix \ref{cha:msgl_determin_macroskopic_parameter}) to yield
\begin{eqnarray}
  \label{eq1:msgl_alpha_beta_fixed}
  \sqrt{\alpha} &=& \pm \frac{2\lambda \cdot  \sqrt{2j \left(1 - \left(\frac{\lambda_{2}}{\lambda}\right)^4\right)}}
  {-i \Gamma + 2\omega} \,;
  \notag \\
  \beta &=& j \left(1 - \left(\frac{\lambda_{2}}{\lambda}\right)^2\right),
\end{eqnarray}
with\cite{Bhaseen_dynamics_of_nonequilibrium_dicke_models}
\begin{eqnarray}
  \label{eq1:lambda_cr_2}
  \lambda_{2}^2 &= \frac{\left(\Gamma^2 + 4\omega^2\right)\omega_0}{16\omega}
  ,
\end{eqnarray}
which differs from the isolated case by the inclusion of the loss-rate $\Gamma$.
 For $\lambda > \lambda_{2}$, $\alpha$ and $\beta$ are non-zero. The coupling $\lambda_{2}$ is our first of three critical points. It is the coupling above which the macroscopic excitation exists. There is also one trivial solution, $\alpha = \beta = 0$, that recovers the normal phase.

\subsection{Analysis of the non-dissipative part of the master equation}
Some insight into the behaviour of the system can be obtained by just considering the non-dissipative part, i.e. the Hamiltonians $H_{N}$ and $H_{S}^{(2)}$ with displacements chosen as in the foregoing.  We study here the eigenvalues of these two Hamiltonians and in the next section we will rewrite  the master equation, Eq. \eqref{eq1:msgl_nagy_form}, in corresponding diagonal bases.

Both Hamiltonians can be diagonalised by a transformation of the following form
\cite{Chaos_and_quantum_phase_Dicke_modell_Clive_Brandes}
 \begin{equation}\label{eq1:diagonal_trafo}
\bf{v} =  \left(
    \begin{array}{cccc}
      \bar{A} & \bar{B} & \bar{G} & \bar{D} \\
      \bar{B} & \bar{A} & \bar{D} & \bar{G} \\
      \bar{A}_2 & \bar{B}_2 &\bar{G}_2 & \bar{D}_2 \\
      \bar{B}_2 & \bar{A}_2 & \bar{D}_2 & \bar{G}_2 \\
    \end{array}
  \right) \cdot
  \left(
    \begin{array}{c}
      d_1 \\
      d_1^\dag \\
      d_2 \\
      d_2^\dag \\
    \end{array}
  \right)\,,
\end{equation}
where
\begin{equation}\bf{v} = \left\{
            \begin{array}{ll}
              (a_1,a_1^\dag,a_2,a_2^\dag)^T, & \hbox{normal phase,} \\
              (c,c^\dag, d,d^\dag)^T, & \hbox{superradiant phase}
            \end{array}
          \right.
 \end{equation}
represents the old basis and $d_i$, the new basis, in which the system is diagonal.
In both cases, the diagonalising matrix has the same structure but different internal parameters, which are listed in Appendix Eq. \eqref{eqa:abgdbar_parameter}-\eqref{eqa:gamma_H02_SR}.

After diagonalisation, $H_{N}$ or $H_{S}^{(2)}$ assume the form $H_D = \overline{\varepsilon}_- d_1^\dag d_1 + \overline{\varepsilon}_+ d_2^\dag d_2$ with eigenenergies
\begin{equation}
\overline{\varepsilon}_{\pm} = \left\{
                                \begin{array}{ll}
                                  \overline{\varepsilon}_{\pm}^{(1)}, & \hbox{ in the normal phase, } \\
                                  \overline{\varepsilon}_{\pm}^{(2)}, & \hbox{ in the superradiant phase}
                                \end{array}
                              \right.
\end{equation}
the forms of which are given in Eqs. \eqref{eqa:energy_H0_NR},\eqref{eqa:energy_H02_SR}.

In the normal phase, one of the eigenenergies $\overline{\varepsilon}_{\pm}^{(1)}$ of $H_{N}$ has a zero at $\lambda = \lambda_{1}$ with
\begin{equation}\label{eq1:lambda_cr_1}
 \lambda_{1}^2  \equiv \frac{\omega \omega_0}{4}\,.
\end{equation}
The vanishing eigenenergy marks the end of the normal phase \cite{Chaos_and_quantum_phase_Dicke_modell_Clive_Brandes}.

\begin{figure}
\begin{center}
  \includegraphics[scale=0.7]{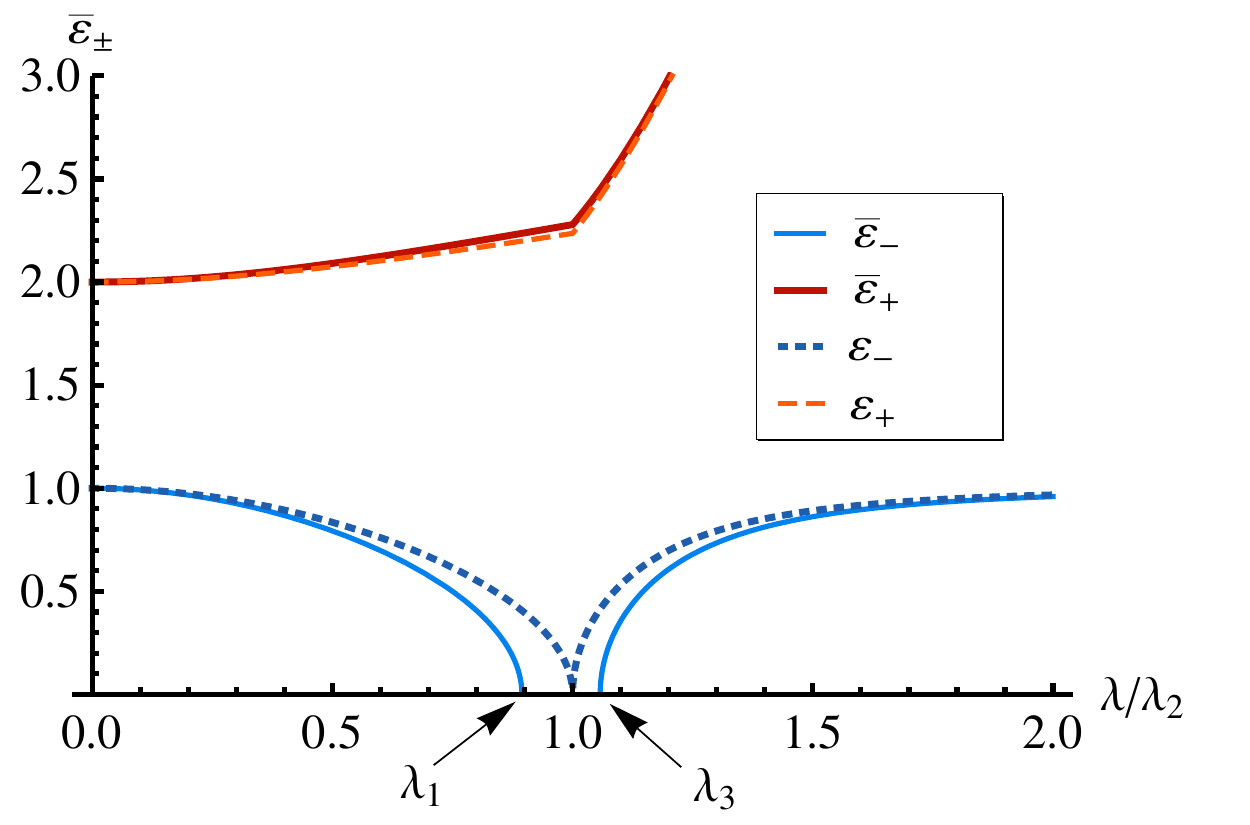}
  \caption
  {(Color online) Eigenenergies of the effective non-equilibrium Dicke-Hamiltonian from the master equation plotted for both phases as a function of $\lambda/\lambda_{2}$ (solid lines).  Also plotted are the energies for the equilibrium ($\Gamma =0$) case (dotted lines).
  For  $\lambda<\lambda_{1}$ we have a normal phase, and for $\lambda>\lambda_{3}$ a superradiant phase. At $\lambda = \lambda_{1}$ and $\lambda = \lambda_{3}$  the $\overline{\varepsilon}_-$ energy vanishes.
  For $\Gamma =0$,  $\lambda_{1}$ and  $\lambda_{3}$ coincide, but out of equilibrium, a gap appears.
  Parameters: $\omega_0 = 2\Gamma, \omega = \Gamma$ (case $\Gamma = 0$: $\omega_0 = 2 \omega$).}
  \label{fig1:Eigenenergien_H_eff_aus_nagy_msc_mit_gap}
\end{center}
\end{figure}

Coming from the superradiant phase, the same eigenenergy vanishes, but in this case at a critical value of $\lambda = \lambda_3$:
\begin{equation}\label{eq1:lambda_cr_3}
 \lambda_{3}^2 =  \frac{\left(\Gamma ^2+4 \omega ^2\right)^{3/2} \omega_0}{16 \cdot 2 \omega^2} \approx
                        \lambda_{1}^2  +\Gamma^2 \frac{3  \omega_0}{32 \omega }+ O(\Gamma^4)\,.
\end{equation}
We recognize that $\lambda_{3}$ and the eigenenergies in the superradiant phase also depend on the bath-coupling-parameter $\Gamma$. Furthermore $\lambda_{1} < \lambda_{3}$, so we obtain a gap, where the eigenenergies become complex and our analysis breaks down, see the discussion below. In  Fig. \ref{fig1:Eigenenergien_H_eff_aus_nagy_msc_mit_gap}, the excitation energy of the closed system (case $\Gamma = 0$) and open system ($\Gamma \neq 0$) are plotted.

\section{Master equation in P-Representation with counting field}
\label{ch:mseq_p_represent_counting}
In the following, our aim is to count how many photons are lost to the environment, and to determine the properties of the photon distribution function. To this end, we introduce a counting field \cite{nazarov2003quantum} $\chi$ into the master equation \eqref{eq1:msgl_nagy_form}
\begin{eqnarray}\label{eq1:msgl_nagy_form_mit_chi}
   \frac{d}{dt} \rho(\chi,t) &=& -i[H_{N/S},\rho(\chi,t)]
   \notag \\
        &&
        \!\!\!\!\!\!\!\!\!\!\!\!\!\!\! \!\!\!\!\!\!\!\!\!\!\!\!\!\!\!
         - \frac{\Gamma}{2} \Biggl{[}
       a_1^\dag  a_1 \rho(\chi,t)  -  2 e^{i\chi} a_1 \rho(\chi,t) a_1^\dag + \rho(\chi,t)  a_1^\dag a_1
    \Biggr{]}
    \!\!.
 \end{eqnarray}
To make our calculations analytically tractable, we make a rotating wave approximation (RWA). We transform the master equation, Eq. \eqref{eq1:msgl_nagy_form_mit_chi},  to a diagonal basis $d_i$ of $H_{N}$ in the normal phase or of $H_{S}^{(2)}$ in the superradiant phase using the relations
\begin{align}\label{eq1:zusammenhang_alte_diag_basis}
    a_1 &= \bar{A} d_1 + \bar{B} d_1^\dag + \bar{G} d_2 + \bar{D} d_2^\dag + \sqrt{\alpha} \,, \notag\\
    a_1^\dag &= \bar{B} d_1 + \bar{A} d_1^\dag + \bar{D} d_2 + \bar{G} d_2^\dag + \sqrt{\alpha^*} \,,
\end{align}
where the coefficients are defined in Appendix \ref{cha:diagonalisation_parameter}.
Next, we transform into the interaction picture and neglect all fast rotating terms such as $d_i d_j^{\dag}$ for $i \neq j$. We obtain the following RWA-master equation for $\rho = \rho(\chi,t)$
\begin{align} \label{eq1:msgl2_rwa_energie_nicht_null}
\dot{\rho} &=  - \frac{\Gamma}{2} \biggl{(}
    \bar{A}^2 (d_1^\dag d_1 \rho - 2 e^{i \chi} d_1 \rho d_1^\dag + \rho d_1^\dag d_1) \notag\\
   &\,\quad\quad + \bar{B}^2  (d_1 d_1^\dag \rho - 2 e^{i \chi} d_1^\dag \rho d_1 + \rho d_1 d_1^\dag) \notag\\
&  \,\quad\quad + \bar{D}^2 (d_2 d_2^\dag \rho - 2 e^{i \chi} d_2^\dag \rho d_2 + \rho d_2 d_2^\dag) \notag\\
 &\,\quad\quad + \bar{G}^2 (d_2^\dag d_2 \rho - 2 e^{i \chi} d_2 \rho d_2^\dag + \rho d_2^\dag d_2)   \biggr{)} \notag \\
 &\quad - \Gamma \abs{\alpha} (1- e^{i \chi}) \rho\,.
\end{align}
This master equation describes two uncoupled harmonic oscillators in interaction with two independent thermal baths at different temperatures \cite{Scully_Qoptics}. We see that each exchange of  quanta with these baths, either into or out, is associated with the emission a of `physical' photon (i.e. one with annihilation operator $a_1$).   The final term in Eq.~(\ref{eq1:msgl2_rwa_energie_nicht_null}) is only nonzero in the superradiant phase and represents a Poissonian process with rate proportional to a macroscopic excitation $\abs{\alpha}$.

Now we transform the master equation \eqref{eq1:msgl2_rwa_energie_nicht_null} into the $P$-representation using the following ansatz
\begin{equation}\label{eq1:Dichtematrix_p_ansatz}
    \rho = \int P(\gamma_1,\gamma_1^*,\gamma_2,\gamma_2^*,\chi,t)\ketbra{\gamma_1}{\gamma_1} \otimes \ketbra{\gamma_2}{\gamma_2}d^2\gamma_1 d^2\gamma_2 \,,
\end{equation}
where $\ket{\gamma_i}$ is the eigenstate of the annihilating operator $d_i$ with eigenvalue $\gamma_i$. Using $d_i \ket{\gamma_i} = \gamma_i \ket{\gamma_i}, d_i^\dag\ket{\gamma_i} = (\pd{}{\gamma_i} + \gamma_i^*) \ket{\gamma_i} $ and integration by parts, we get a partial differential equation for the quasi distribution $P=P(\gamma_1,\gamma_1^*,\gamma_2,\gamma_2^*,\chi,t)$,

\begin{align}\label{eq1:msgl_chi_rwa_p_dgl}
    \dot{P} &= \hat{O} P  - \Gamma  \abs{\alpha} (1 - e^{i\chi}) P\,, \\
\intertext{with}
      \hat{O} &= \bar{U} \left(\pd{}{\gamma_1} \gamma_1 + \pd{}{\gamma_1^*} \gamma_1^*\right)
            + \bar{V} \left(\pd{}{\gamma_2} \gamma_2 + \pd{}{\gamma_2^*} \gamma_2^*\right) \notag \\
             &\quad  + \left(\bar{W} \pd{}{\gamma_1} \pd{}{\gamma_1^*} + \bar{T} \pd{}{\gamma_2} \pd{}{\gamma_2^*}\right)
             + \Gamma (1-e^{i\chi}) \times \notag\\
            & \quad \quad  \left( (\bar{A}^2 + \bar{B}^2) \gamma_1 \gamma_1^* + (\bar{G}^2 + \bar{D}^2) \gamma_2 \gamma_2^* + \bar{B}^2 +  \bar{D}^2\right) \mathbb{1} ;\notag\\
\bar{U} &= \frac{\Gamma}{2} \left(\bar{A}^2 + \bar{B}^2\left(1- 2 e^{i \chi}\right)\right) ; \notag\\
\bar{V} &= \frac{\Gamma}{2}  \left(\bar{G}^2 + \bar{D}^2\left(1- 2 e^{i \chi}\right)\right) ;  \notag\\
\bar{W} &= \Gamma \bar{B}^2 e^{i\chi} ; \notag\\
\bar{T} &= \Gamma \bar{D}^2 e^{i\chi}\,.
\end{align}
Equation \eqref{eq1:msgl_chi_rwa_p_dgl} can be solved as
\begin{equation}\label{eq1:p_dgl_lsg_asnatz_1}
  P = P_0(t) \cdot e^{-\Gamma \abs{\alpha} (1- e^{i \chi})t}\,,
\end{equation}
if
\begin{equation}\label{eq1:p_dgl_2}
  \dot{P}_0(t) = \hat{O} P_0(t)
\end{equation}
is fulfilled. Since $\hat{O}$ is a bi-linear operator, the solution of Eq. \eqref{eq1:p_dgl_2} can be written in the following form
\begin{align}\label{eq1:msgl_rwa_p_dgl_loesung_ansatz}
    P_0 &= \exp \bigl{(}-a(\chi,t) + b_1(\chi,t)\cdot \gamma_1 + b_2(\chi,t)\cdot \gamma_2 \notag \\
            &\quad \quad \quad + c_1(\chi,t) \cdot \gamma_1^* +  c_2(\chi,t) \cdot \gamma_2^* \notag\\
            &\quad \quad \quad- d_1(\chi,t) \cdot \gamma_1 \gamma_1^* - d_2(\chi,t) \cdot \gamma_2 \gamma_2^* \bigr{)}\,.
\end{align}
Substituting \eqref{eq1:msgl_rwa_p_dgl_loesung_ansatz} into  $\dot{P}_0 = \hat{O} P_0$ we obtain seven coupled first order differential equations for the functions $a, b_i, c_i, d_i$, see Appendix Eq. \eqref{eqa:msgl2_chi_rwa_p_dglsystem}. Taking a displaced 2-dimensional gaussian distribution with a standard deviation $\varepsilon$ as an initial condition,
\begin{equation}\label{eq1:msgl_rwa_p_dgl_ab}
    P_0(\gamma_1,\gamma_1^*,\gamma_2,\gamma_2^*,\chi,t=0) = \frac{1}{(2 \pi \varepsilon)^2}
                        \cdot e^{\frac{-|\gamma_1 - \gamma_1^0|^2 - |\gamma_2 - \gamma_2^0|^2}{\varepsilon}}\,,
\end{equation}
we can solve this system analytically. The steady state solution for $t\to \infty$ can also be calculated and is listed in Appendix Eq. \eqref{eqa:msgl2_chi_rwa_p_dglsys_lsg_t_gross}.

With knowledge of the $P$-representation of the density matrix with the counting field $\chi$, we can now calculate the mean occupation of the original modes and the cumulant generating function for the photon counting statistics.

As a final remark, the RWA would normally not be valid exactly near the points $\lambda_{1}, \lambda_2$, because here one of the eigenenergies $\varepsilon_{\pm}$ vanishes.  Terms like $\tilde{d}_1^{(\dag)} \tilde{d}_1^{(\dag)}$ can no longer be neglected on the basis that they are quickly rotating.

However, we repeated our calculations ($\chi=0$) in the singular coupling limit of the master equation where terms like $e^{\pm i \varepsilon_{\pm}t}$ approach unity and found that, for the parameters studied, no significant qualitative differences with the RWA method arose.

\section{Results}
\label{ch:results}
\subsection{System properties}
We first calculate the occupation number of the photonic and atomic modes. In our $P$-representation, an operator average is given by
\begin{equation}\label{eq1:Erwartungswert_mit_p}
\avg{\hat{A}} = \text{Tr}{\hat{A} \rho} = \text{Tr} \int \hat{A} P \,\ketbra{\gamma_1}{\gamma_1} \otimes \ketbra{\gamma_2}{\gamma_2} \, d\gamma_1^2 d\gamma_2^2\,,
\end{equation}
which we use to calculate $\avg{a_1^\dag a_1}$ and $\avg{a_2^\dag a_2}$. We have to use the relation between the old and the diagonal basis, because the $P$-representation has been evaluated in the diagonal basis. We also have to do the calculation in the interaction picture and use the RWA, as we have done in the calculation of the $P$-function. In both phases the time-dependent average of the optical and atomic modes has the same structure of the form $\avg{a_i^\dag a_i}(t) = e^{-\Gamma \cos^2\gamma t} (...) + e^{-\Gamma \sin^2\gamma t} (...) + \avg{a_i^\dag a_i}(t\to\infty)$ from which we can read off the characteristic relaxation times
\begin{equation}\label{eq1:relax_time}
\tau_1 \equiv (\Gamma \cos^2\gamma)^{-1} \text{ and } \tau_2 \equiv (\Gamma \sin^2\gamma)^{-1}\,,
\end{equation}
where $\gamma$ is the rotation angle of the decoupling, Eqs. \eqref{eqa:gamma_H0_NR},\eqref{eqa:gamma_H02_SR}.
As a result, the two modes develop at two different time scales. In the following, we look at the long-time solution.

In the normal phase, using the long-time solution of Eq. \eqref{eq1:msgl_chi_rwa_p_dgl} we get (see Appendix \ref{cha:avarage_and_p})
\begin{align}\label{eq1:moden_avg_nr}
    \avg{a_1^\dag a_1}(t\to\infty) &= \frac{\lambda^2}{2\omega \omega_0 - 8 \lambda^2}; \notag \\
    \avg{a_2^\dag a_2}(t\to\infty) &= \frac{1}{8}
                    \left(-4+\frac{\omega}{\omega_0}+\frac{2 \omega_0}{\omega}+\frac{\omega ^2}{-4 \lambda ^2+\omega  \omega_0}\right) \,,
\end{align}
from which we recognize the expected divergence at $\lambda = \lambda_{1}$, Eq. \eqref{eq1:lambda_cr_1}.

In the superradiant phase, we have a finite macroscopic occupation, which is much greater that the fluctuations around it. We obtain
\begin{align} \label{eq1:moden_avg_sr}
\avg{a_1^\dag a_1} &= \avg{c^\dag c} + \abs{\alpha}; \notag  \\
\avg{a_2^\dag a_2} &= \avg{d^\dag d} + \abs{\beta}\,, \\
\intertext{where in the stationary case $t \to \infty$}
  \avg{c^\dag c} &= \frac{1}{8} \biggl{(}-2+ \frac{\omega ^2 \cos^2(\gamma)}{\left(\overline{\varepsilon} _-^{(2)}\right)^2}+\frac{\cos^2(\gamma) \left(\overline{\varepsilon} _-^{(2)}\right)^2}{\omega ^2} \notag\\
  &+\frac{\omega ^2 \sin^2(\gamma)}{\left(\overline{\varepsilon} _+^{(2)}\right)^2}+\frac{\sin^2(\gamma) \left(\overline{\varepsilon} _+^{(2)}\right)^2}{\omega ^2}\biggr{)} ; \\
  \avg{d^\dag d} &= \frac{1}{8} \biggl{(}-4+\frac{\omega }{\Omega_0}+\frac{\Omega_0}{\omega }+\frac{\omega  \Omega_0 \sin^2(\gamma)}{\left(\overline{\varepsilon} _-^{(2)}\right)^2} \notag\\
  & \quad +\frac{\sin^2(\gamma) \left(\overline{\varepsilon} _-^{(2)}\right)^2}{\omega  \Omega_0}+\frac{\omega  \Omega_0 \cos^2(\gamma)}{\left(\overline{\varepsilon} _+^{(2)}\right)^2} \notag \\
  &\quad +\frac{\cos^2(\gamma) \left(\overline{\varepsilon} _+^{(2)}\right)^2}{\omega  \Omega_0}\biggr{)}\,.
\end{align}

\begin{figure}
\center
  \includegraphics[scale=0.7]{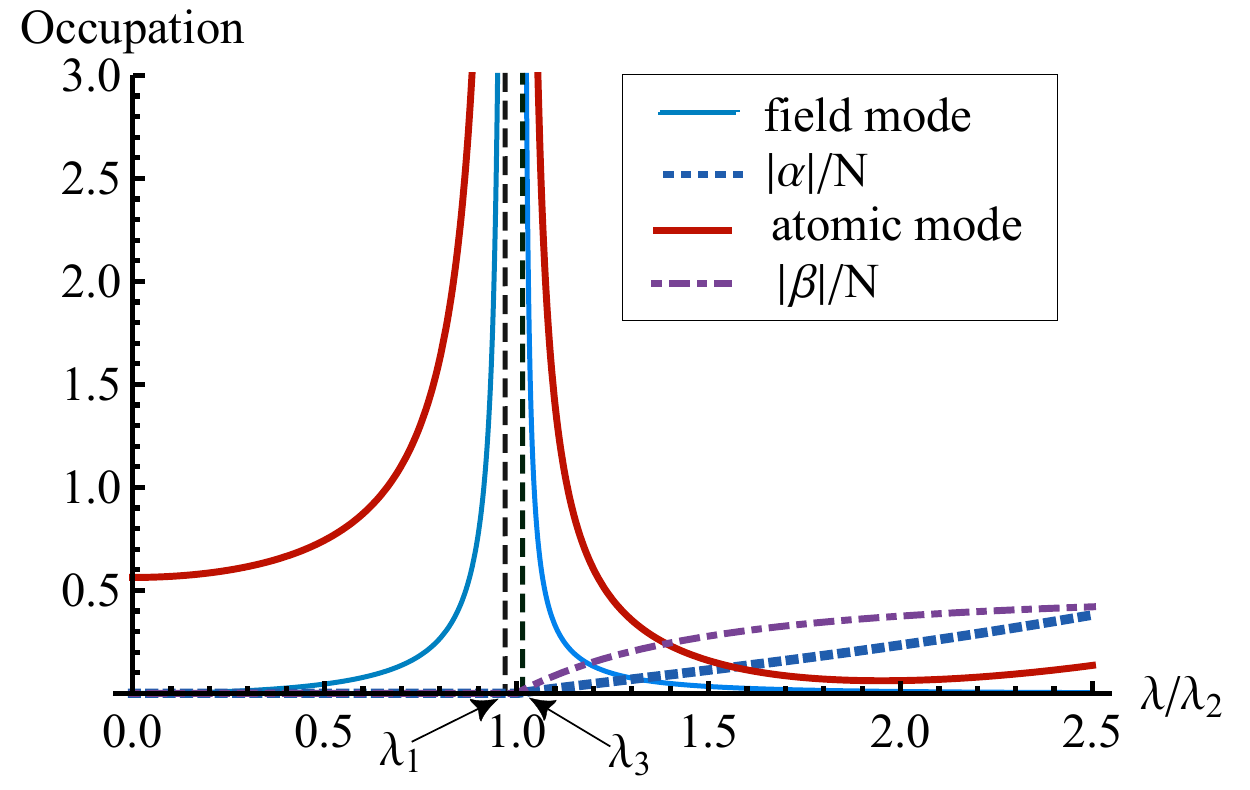}\\
  \caption{(Color online) Fluctuation of the field and atomic mode occupation number (solid line), Eq. \eqref{eq1:moden_avg_nr},\eqref{eq1:moden_avg_sr}  and the correspondent macroscopic occupation  (dashed line). Parameters:  $\omega = 2\Gamma,\omega_0=0.5\Gamma$.}\label{fig1:msgl2_rwa_NPSR_stat_besetz_v3}
\end{figure}

Fig. \ref{fig1:msgl2_rwa_NPSR_stat_besetz_v3} shows the occupation number of the field and the atomic modes as a function of $\lambda$ in the steady state and gives results that are approximately equivalent to those of Nagy and co-workers \cite{Nagy_critical_exponent_open_dicke_model}. We see that the fluctuations diverge around the phase transition, but there is again the undefined area for $\lambda \in [\lambda_{1},\lambda_{3}]$. This area has been also seen in the energy plot.

In the normal phase the occupation of both modes decrease with smaller coupling parameter $\lambda$. Decreasing $\lambda$ weakens the coupling between the optical and the field mode. For $\lambda \to 0$ the atomic mode decouples completely and only the field mode becomes damped.

In the superradiant phase, the occupation of both modes first decreases with increasing $\lambda$ but then the occupation of the field mode increases. The increasing $\lambda$ reduces the coupling $\Lambda$, Eq. \eqref{eq1:H2-parameter}, which explains the decreasing. On the other hand, the $M$-term in the Hamiltonian then becomes dominant. For $\lambda \gg \lambda_{3}$ the $d$-mode becomes approximately  decoupled but squeezed because of the presence of the $M$-term, see Eq. \eqref{eq1:H0_SR_mit H02}, which explains the increase of occupation. That means, only the $c$-mode is then damped by the bath.

The macroscopic occupation $\abs{\alpha}, \abs{\beta}$ exists only in the superradiant phase, where it is much greater than the fluctuation, and  vanishes for $\lambda < \lambda_{2}$.

\subsection{Counting statistics}
We now derive the cumulant generating function (CGF) $F(\chi,t)$ of the photon counting statistics\cite{nazarov-FCS_charge_transfer_coulomb_blockade}, defined as
\begin{equation}\label{eq1:F_def_original}
  F(\chi,t)
  = \log \text{Tr} \rho(\chi,t)\,
  .
\end{equation}

The $k$-th cumulant can be obtained by
\begin{equation}\label{eq1:kumm_aus_F_def}
\avg{n^k}_C = \left.\frac{\partial^k}{\partial (i\chi)^k} F(\chi,t) \right|_{\chi=0}\,.
\end{equation}
In the long-time limit,  the CGF can be calculated from our $P$-function including a counting field $\chi$, Eq. \eqref{eq1:msgl_rwa_p_dgl_loesung_ansatz} as follows
\begin{align}\label{eq1:kumm_generating_fnc}
F(\chi,t\to \infty)&=  \log\int P(o_1,o_1^*,o_2,o_2^*,\chi,t \to \infty) d^2o_1 d^2o_2 \notag \\
        &= -\Gamma \abs{\alpha} (1- e^{i\chi})t \notag \\
        & \quad \quad + \log \left(\frac{4 \pi^2}{d_1(\chi,t) d_2(\chi,t)} \cdot e^{-a(\chi,t\to \infty)}
                \right) \notag\\
        &\approx -\Gamma \abs{\alpha} (1- e^{i\chi})t  - a(\chi,t \to \infty),
\end{align}
where $a(t \to \infty,\chi)$ is defined in Appendix Eq.  \eqref{eqa:msgl2_chi_rwa_p_dglsys_lsg_t_gross} and terms $d_i \ll a$ have being neglected. Inserting the quantity $a(\chi, t \to \infty)$ from Eq. \eqref{eqa:msgl2_chi_rwa_p_dglsys_lsg_t_gross}, we obtain the CGF for both phases,
 \begin{align}\label{eq1:kum_generating_fnc_endergebnis}
\lim_{t\to \infty} \frac{1}{t} F(\chi,t) &=   \Gamma \abs{\alpha} (e^{i\chi}-1) + \frac{1}{2} \biggl{(}\bar{A}^2-\bar{B}^2 + \bar{G}^2 -\bar{D}^2 \notag \\
&\quad-\sqrt{\bar{A}^4+\bar{B}^4-2 \bar{A}^2 \bar{B}^2 \left(2 e^{2 i \chi }-1\right)} \notag \\
&\quad -\sqrt{\bar{D}^4 + \bar{G}^4-2 \bar{D}^2 \bar{G}^2 \left(2 e^{2 i \chi }-1\right)}\biggr{)} \Gamma \,.
\end{align}
The CGF thus consists of two parts. The first is the CGF of a Poissonian process with rate $\Gamma |\alpha|$. In the superradiant phase, $|\alpha|$ is macroscopic (proportional to the number of atoms) and this contribution dominates.  The k-th cumulants from this macroscopic contribution are simply $\avg{n^k}_C = \Gamma\abs{\alpha}$. On top of this contribution, which is absent in the normal phase (where $\alpha=0$), comes a further contribution to the CGF which arises from fluctuations about the mean-field displacements.  The cumulants due to this contribution are given by $\delta \avg{n^k}_C
  = - \frac{\partial^k}{\partial (i\chi)^k} a(\chi,t)|_{\chi=0}$.

With this result we are now able to calculate all cumulants in the long-time limit. For example, the first cumulant reads
\begin{equation} \label{eq1:erste_kumm_ergebnis}
\avg{n}_C = \abs{\alpha} \Gamma t + \delta \avg{n}_C\,,
\end{equation}
where
\begin{equation}
\delta \avg{n}_C = \Gamma  t \cdot \left\{
                                                 \begin{array}{ll}
                                                   \avg{a_1^\dag a_1}, & \hbox{normal phase} \\
                                                   \avg{c^\dag c}, & \hbox{superradiant phase.}
                                                 \end{array}
                                               \right.
\end{equation}
We see that $\delta\avg{n}_C$ is proportional to the mean occupation of the optical mode that directly couples to the environment. This fits with the experiment use of the mean macroscopic flux $\sim \abs{\alpha}$ as a measure of the order parameter of the superradiant phase transition \cite{Baumann-Dicke_qpt,ritter_brennecke_dynamical_coupling_between_bec_optical_lattice}.  As must be, the cumulant generating function is linear in time. We mention that the source \footnote{Setting $\Omega = 0$ and deriving the master equation using $H$ and $H_1$ again gives a master equation for $\rho$ similar to Eq. \eqref{eq1:msgl2_rwa_energie_nicht_null} but without let the 'heating terms' $\sim \bar{B},\bar{D}$. } of the photon current is the driving frequency $\Omega$.
\begin{figure}
\center
\includegraphics[scale=0.7]{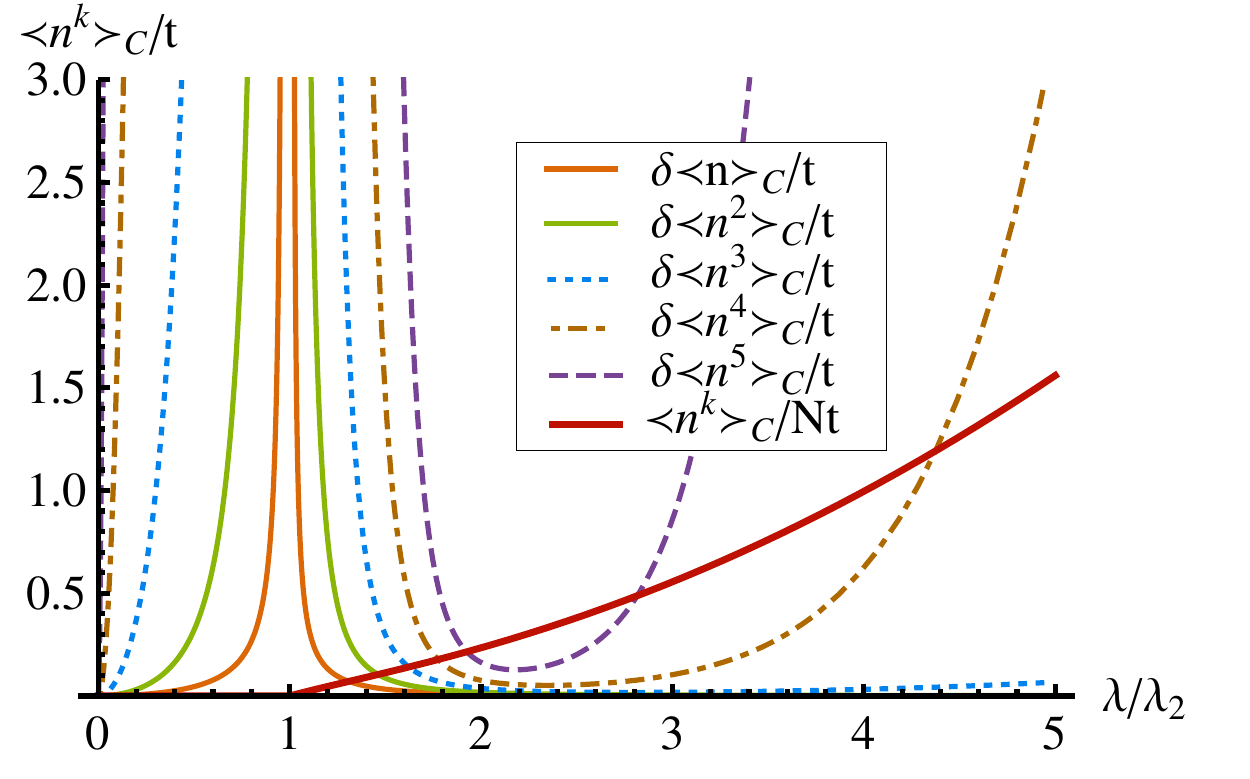}
\caption{(Color online) Asymptotic cumulants divided by time in both phases for driven dissipative Dicke model. $\avg{n^k}_C \sim N$ is macroscopic, $\delta \avg{n^k}_C$ is the corresponding fluctuation, Eq. \eqref{eq1:erste_kumm_ergebnis}. For $\lambda \gg \lambda_{2}$ the cumulants starting by the third are growing. We can see also a tiny gap around $\lambda_{2}$ corresponding to the undefined area $\lambda_1 < \lambda_3$ of our theory. Parameters: $\omega = 2 \Gamma, \omega_0 = 0.5 \Gamma $.
}
\label{fig1:msgl2_rwa_kumsteigung_fano_gegen_l}
\end{figure}
\begin{figure}
\includegraphics[scale=0.7]{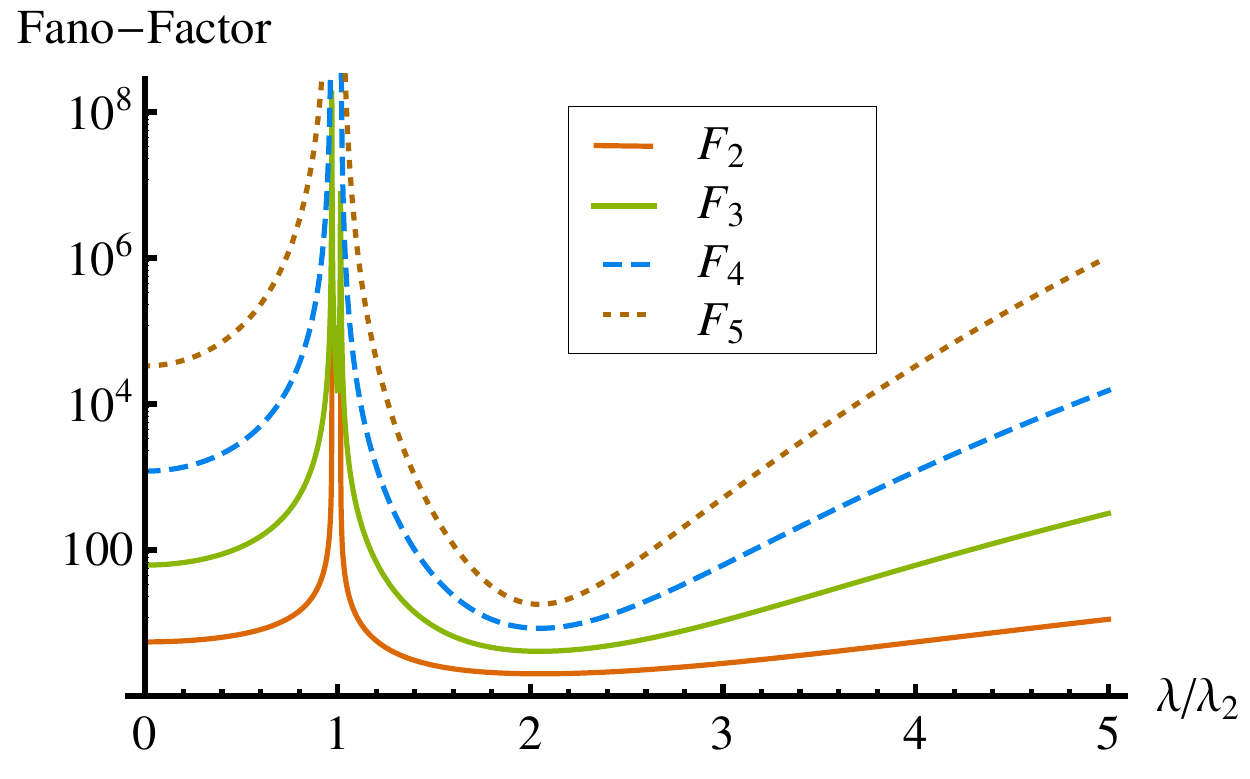}
\caption{(Color online) Same as Fig. \ref{fig1:msgl2_rwa_kumsteigung_fano_gegen_l} but with Fano factors not cumulants.
\label{fig1:msgl2_rwa_fano_gegen_l}
}
\end{figure}

In Fig. \ref{fig1:msgl2_rwa_kumsteigung_fano_gegen_l} we have plotted the first five cumulants divided by time as a function of the coupling parameter $\lambda$. We see all cumulants diverging near the phase transition $\lambda \approx \lambda_2$, where one eigenvalue of the Hamiltonian becomes zero. Going away from the critical region, our system emits less to the environment and therefore $\delta\avg{n^k}/t$ decreases. For $\lambda \gg \lambda_{3}$, $\delta\avg{n}_C, \delta\avg{n^2}_C$ become constant, however, higher cumulants such as $\delta\avg{n^3}_C$ grow. However, in the superradiant phase all fluctuations can be neglected because of the existence of a macroscopic Poisson distribution $\avg{n^k}_C$ (Note the scaling $1/N$ in Fig. \ref{fig1:msgl2_rwa_kumsteigung_fano_gegen_l}), which is much greater than the fluctuation.  We see again the undefined area around $\lambda_{2}$. Fig. \ref{fig1:msgl2_rwa_fano_gegen_l} shows the fluctuation part of the Fano factors $F_i \equiv \frac{\delta\avg{n^i}_C}{\delta\avg{n}_C}$.

\begin{figure}
\begin{center}
  \includegraphics[scale=0.7]{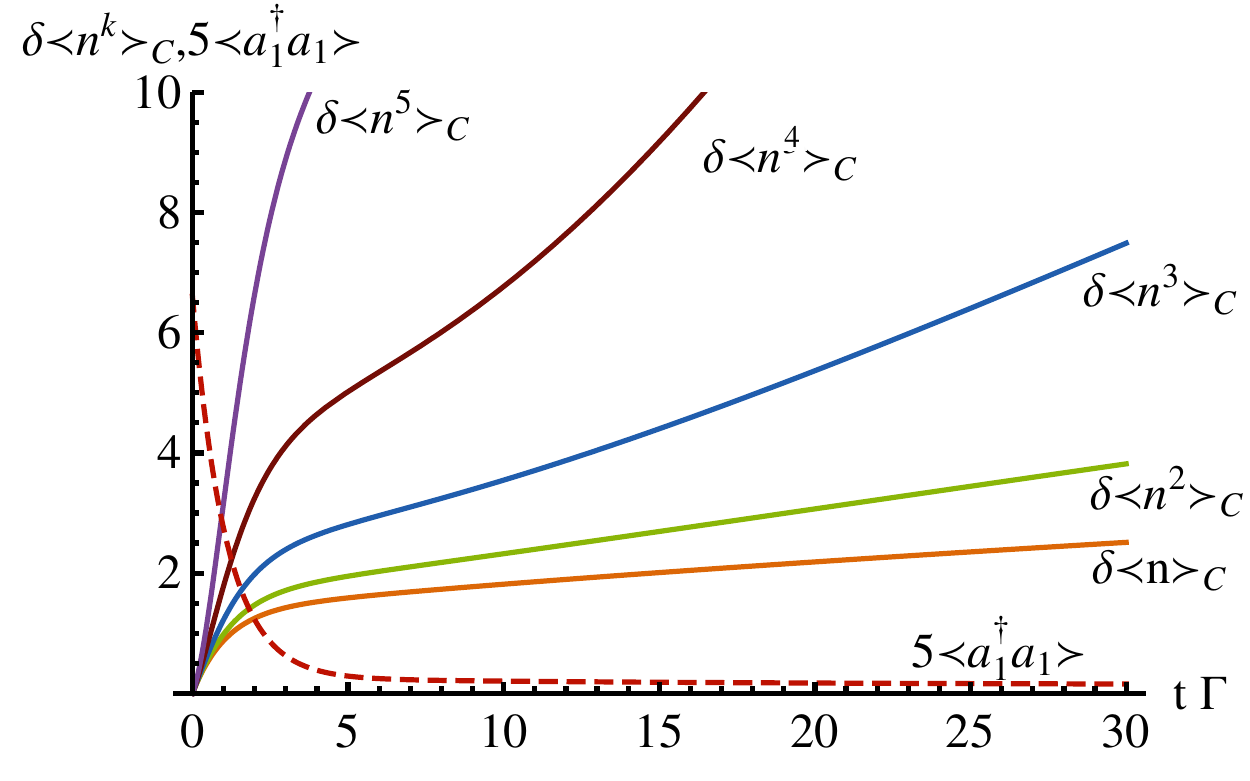}
  \caption{(Color online) Temporal development of the first five cumulants and of the photon mode occupation average $\avg{a_1^\dag a_1}$. After some characteristic time scale, $\avg{a_1^\dag a_1}$ approximately reaches the steady state value, and on the same time scale the cumulants change their slope. Parameters: $\omega_0 = 2\Gamma, \omega = 1\Gamma, \lambda = 0.3 /\Gamma$.}
  \label{fig1:kumulant_time_development}
\end{center}
\end{figure}

If we use the complete solution of Eq.  \eqref{eq1:msgl_chi_rwa_p_dgl}, we can calculate the cumulants as a function of time $t$. The evolution of the first cumulants is shown in Fig. \ref{fig1:kumulant_time_development} . We see that there are at least two time scales. This behaviour is due to the two timescales $\tau_1$ and $\tau_2$, Eq. \eqref{eq1:relax_time}, of the both modes. As we can see in the same picture, $\avg{a_1^\dag a_1}$ develops in an analogous fashion.

\section{Discussion}
\label{ch:discussion}
We have derived the cumulant generating function for a driven single-mode Dicke Hamiltonian coupled to a bath.  The combination of the thermodynamic limit and the use of the $P$ representation allows us to obtain exact expressions not only for the asymptotic CGF and cumulants, but also those at finite times.

The CGF and photon-counting cumulants consist of two parts: a macroscopic (order $N^1$) contribution and fluctuations around it (order $N^0$).  Although the fluctuation component would be masked by the macroscopic contribution,  this latter is only non-zero in the superradiant phase, such that the fluctuation-component could be experimentally accessed in the normal phase.

We have identified three critical values $\lambda_1,\lambda_2,\lambda_3$ for the coupling constant $\lambda$. The value $\lambda_2$, Eq. \eqref{eq1:lambda_cr_2}, marks the phase transition, the area $\lambda > \lambda_{2}$ has a non-vanishing macroscopic mode occupation $\abs{\alpha},\abs{\beta}$, Eq. \eqref{eq1:msgl_alpha_beta_fixed}.
Both the value of $\lambda_2$ and of $\abs{\alpha},\abs{\beta}$ coincide with results in the literature \cite{Dimer-Carmichael_Proposed_realization_of_dicke_model_qpt_optical_cavity,oztop-excitations_of_otpically_driven_atomic_condensata_theory_photodetection,Bhaseen_dynamics_of_nonequilibrium_dicke_models}. The expression for $\lambda_{1}$ and $\lambda_{3}$ also agree with previous calculations based an equations of motions \cite{Dimer-Carmichael_Proposed_realization_of_dicke_model_qpt_optical_cavity,oztop-excitations_of_otpically_driven_atomic_condensata_theory_photodetection}, however, only up to the first order in $\Gamma$ (which is a small correction anyway). For example, our result \footnote{We include the dissipative dynamics in the effective Hamiltonian by rewriting the master equation, Eq. \eqref{eq1:msgl_nagy_form}, to $\frac{d}{dt} \rho = - i [H_{\text{eff}},\rho(t)] + \Gamma a_1 \rho a_1^\dag$ with non-hermitian $H_{\text{eff}} = H - i\frac{\Gamma}{2} a_1^\dag a_1$. One of its eigenvalues has a zero at $\lambda'_1 = \frac{\sqrt{\omega - i \frac{\Gamma}{2}}\sqrt{\omega_0}}{2}$.  Its real part for $\omega = \omega_0$ $\Re (\lambda'_1) = \frac{\omega}{2} + \frac{\Gamma^2}{64 \omega} + O(\Gamma^4)$ corresponds to \text{ \cite{Dimer-Carmichael_Proposed_realization_of_dicke_model_qpt_optical_cavity}}, but only up to a factor 2 in the $\Gamma^2$ term.}
for $\lambda_1$,$\lambda_3$, Eq. \eqref{eq1:lambda_cr_1}\eqref{eq1:lambda_cr_3} differs from that of Dimer\cite{Dimer-Carmichael_Proposed_realization_of_dicke_model_qpt_optical_cavity} et al. by a factor 2 in the term of $O(\Gamma^2)$.
The mean photon occupation number calculated for the normal phase by {\"O}ztop\cite{oztop-excitations_of_otpically_driven_atomic_condensata_theory_photodetection} et al. coincides with our result up to the first order in $\Gamma$. We ascribe these small difference as to the different diagonalisation procedures for the effective Hamiltonians.

As in previous works\cite{Dimer-Carmichael_Proposed_realization_of_dicke_model_qpt_optical_cavity,oztop-excitations_of_otpically_driven_atomic_condensata_theory_photodetection,Nagy_critical_exponent_open_dicke_model}, the status of the region $\lambda_1 < \lambda < \lambda_3$, around the critical point making the normal-superradiant transition remains an open issue.  In our approach, the effective Hamiltonian is not stable in that region. This could indicate a limitation of the simple mean field approach close to $\lambda_{2}$ in the dissipation model\footnote{But choosing the macroscopic occupation $\alpha,\beta$ in a way to fulfill $H_S^{(1)}=0$  and applying the RWA to the Eq. \eqref{eq1:msgl_nagy_form_mean_field_trafo}, all the critical values $\lambda _{1,2,3}$ coincide. In this case the critical coupling for the phase transition is independent of the bath coupling $\Gamma$}.

\begin{acknowledgments}
We thank I. Lesanovsky and  S. Genway for useful discussions.  The authors gratefully acknowledge financial support from the DAAD and DFG Grants BR $1528/7-1$, $1528/8-1$, SFB $910$, and GRK $1558$.

\end{acknowledgments}



\appendix
\section{Diagonalisation parameter}
\label{cha:diagonalisation_parameter}
In Eq. \eqref{eq1:diagonal_trafo}, the parameters for the diagonalisation are
\begin{align}\label{eqa:abgdbar_parameter}
 \bar{A} &=  \frac{\cos(\gamma) (\omega + \overline{\varepsilon}_-)}{2\sqrt{\omega} \sqrt{\overline{\varepsilon}_-}} ;
& \bar{B} &= \frac{\cos(\gamma) (\omega - \overline{\varepsilon}_-)}{2\sqrt{\omega} \sqrt{\overline{\varepsilon}_-}}; \notag \\
\bar{G}  &= \frac{\sin(\gamma) (\omega + \overline{\varepsilon}_+)}{2\sqrt{\omega} \sqrt{\overline{\varepsilon}_+}} ;
&\bar{D}  &= \frac{\sin(\gamma) (\omega - \overline{\varepsilon}_+)}{2\sqrt{\omega} \sqrt{\overline{\varepsilon}_+}}; \notag \\
\intertext{}
\overline{A}_2 &=  \frac{-\sin(\gamma) (\Omega_0 + \overline{\varepsilon}_-)}{2\sqrt{\Omega_0} \sqrt{\overline{\varepsilon}_-}} ;
& \overline{B}_2 &= \frac{\sin(\gamma) (-\Omega_0 + \overline{\varepsilon}_-)}{2\sqrt{\Omega_0} \sqrt{\overline{\varepsilon}_-}}; \notag \\
\overline{G}_2  &= \frac{\cos(\gamma) (\Omega_0 + \overline{\varepsilon}_+)}{2\sqrt{\Omega_0} \sqrt{\overline{\varepsilon}_+}};
&\overline{D}_2  &= \frac{\cos(\gamma) (\Omega_0 - \overline{\varepsilon}_+)}{2\sqrt{\Omega_0} \sqrt{\overline{\varepsilon}_+}}\,.
\end{align}
In the normal phase, $\omega=\Omega_0$. The eigenenergies are
\begin{align}\label{eqa:energy_H0_NR}
{2\overline{\varepsilon}_{\pm}^{(1)}}^2 &=
\left\{
\begin{array}{ll}
 \omega^2+\omega_0^2
                \pm \sqrt{\left(\omega_0^2 - \omega^2\right)^2 + 16\lambda^2 \omega \omega_0} ,
                & \hbox{} \omega_0 > \omega\\
\omega^2+\omega_0^2
                \mp \sqrt{\left(\omega_0^2 - \omega^2\right)^2 + 16\lambda^2 \omega \omega_0} \,,
                & \hbox{} \omega_0 \leq \omega \,.
\end{array}
\right.
\end{align}
The rotation angle for the decoupling is
\begin{equation}\label{eqa:gamma_H0_NR}
\tan(2\gamma) = \frac{4\lambda \sqrt{\omega \cdot \omega_0}}{\omega_0^2 - \omega^2} \,.
\end{equation}
In the superradiant phase, the eigenenergies of $H_S^{(2)}$ are
\begin{align}\label{eqa:energy_H02_SR}
 2 {\overline{\varepsilon}_{\pm}^{(2)}}^2 &=
    \left\{
      \begin{array}{ll}
        \omega ^2+4 M  \Omega_0+\Omega_0^2 \pm \sqrt{h}, & \hbox{case of } \omega_0 >\omega \\
        \omega ^2+4 M  \Omega_0+\Omega_0^2 \mp \sqrt{h}, & \hbox{case of } \omega_0 \leq \omega \,,
      \end{array}
    \right.    \\
\intertext{with}
   h&= \omega ^4+\Omega_0 \biggl{(}-8 \omega  \left(-2 \Lambda ^2+M  \omega \right) \notag\\
    &\quad +\Omega_0 \left(16 M ^2-2 \omega ^2+8 M  \Omega_0+\Omega_0^2\right)\biggr{)}\,.
\end{align}
The rotation angle fulfills
\begin{align}\label{eqa:gamma_H02_SR}
 -2 h \cos^2 (\gamma) &=  \omega ^2 \left(\sqrt{h}+\omega ^2\right)-\left(\sqrt{h}-16 M^2+2 \omega ^2\right) \Omega _0^2 \notag \\
 & \quad + 4 \left(4 \Lambda ^2 \omega -M \left(\sqrt{h}+2 \omega ^2\right)\right) \Omega _0 \notag\\
& \quad +8 M \Omega _0^3+\Omega _0^4 \,.
\end{align}

\section{Master equation}
\label{cha:msgl_determin_macroskopic_parameter}

The three equations for the macroscopic occupation parameters $\alpha,\alpha^*,\beta$, Eq. \eqref{eq1:msgl_alpha_beta_fixed}, are
\begin{align}
\label{eqa:mscgl_naggy_form_verschieb_festlegen_gl_2_1}
0 &= - 2\lambda \sqrt{\frac{k}{2j}} \sqrt{\beta} + \sqrt{\alpha} (\omega - i \frac{\Gamma}{2}) \notag \\
  &\quad        \Rightarrow  \sqrt{\alpha} = \frac{2 \lambda \sqrt{\frac{k}{2j}} \sqrt{\beta}}{\omega - i \frac{\Gamma}{2}}  ;\\
\label{eqa:mscgl_naggy_form_verschieb_festlegen_gl_2_2}
0&=  - 2\lambda \sqrt{\frac{k}{2j}} \sqrt{\beta} + \sqrt{\alpha^*} (\omega + i \frac{\Gamma}{2}) \notag\\
 &           \Rightarrow \sqrt{\alpha^*} = \frac{2 \lambda \sqrt{\frac{k}{2j}} \sqrt{\beta}}{\omega + i \frac{\Gamma}{2}} ; \\
\label{eqa:mscgl_naggy_form_verschieb_festlegen_gl_2_3}
0&=\lambda \sqrt{\frac{k}{2j}} (\sqrt{\alpha} + \sqrt{\alpha^*})\left(1 - \frac{\beta}{k}
                         \right) - \omega_0 \sqrt{\beta}\,. &
\end{align}
Insertion of \eqref{eqa:mscgl_naggy_form_verschieb_festlegen_gl_2_1},\eqref{eqa:mscgl_naggy_form_verschieb_festlegen_gl_2_2} in \eqref{eqa:mscgl_naggy_form_verschieb_festlegen_gl_2_3} produces
\begin{align}
  0 &=   \lambda \cdot \sqrt{\frac{k}{2j}}
         \left( \frac{2 \lambda \sqrt{\frac{k}{2j}}}{\omega + i \frac{\Gamma}{2}}
                +
                \frac{2 \lambda \sqrt{\frac{k}{2j}}}{\omega - i \frac{\Gamma}{2}}
        \right) \sqrt{\beta} \cdot \left(1 - \frac{\beta}{k}\right) \notag \\
    & \quad - \omega_0 \sqrt{\beta}\,.  \\
\intertext{Using $ 1 - \frac{\beta}{k} =  \frac{2j - 2\beta}{k}$ this expression can be simplified to}
0 &= \left(2(j - \beta)\frac{\lambda^2}{j}
\underbrace{\left(\frac{1}{\omega + i \frac{\Gamma}{2}} + \frac{1}{\omega -i \frac{\Gamma}{2}}\right)}_{=\frac{8 \omega}{\Gamma^2 + 4 \omega^2}}
- \omega_0\right) \cdot \sqrt{\beta} \,, \\
\intertext{so that we get two solutions, $\beta = 0$ or}
 \beta &= j \left( 1- \frac{\Gamma^2 \omega_0}{16 \lambda^2 \omega} - \frac{\omega \omega_0}{4 \lambda^2}\right) = j \left(1 - \left(\frac{\lambda_{2}}{\lambda}\right)^2\right) \,.
\end{align}
Using these values of $\alpha,\beta$, we insert them into the parameters of $H_S^{(2)}$, Eq. \eqref{eq1:H2-parameter},
\begin{align}\label{eqa:msgl_H02_parameter_fixed}
\Lambda &=\frac{\left(\Gamma ^2+4 \omega ^2\right) \omega_0}{2 \sqrt{2} \sqrt{\omega  \left(16 \lambda ^2 \omega +\left(\Gamma ^2+4 \omega ^2\right) \omega_0\right)}} ; \\
\Omega_0&=\frac{8 \lambda ^2 \omega }{\Gamma ^2+4 \omega ^2}+\frac{\omega_0}{2}; \notag \\
M&=-\frac{\omega_0}{8}+\frac{96 \lambda ^4 \omega ^2-2 \lambda ^2 \omega  \left(\Gamma ^2+4 \omega ^2\right) \omega_0}{\left(\Gamma ^2+4 \omega ^2\right) \left(\Gamma ^2 \omega_0+4 \omega  \left(4 \lambda ^2+\omega  \omega_0\right)\right)}\,. \notag
\end{align}
Finally, the set of coupled differential equations for the $P$-representation, Eq. \eqref{eq1:p_dgl_2}, reads (functions depends on $t$ and $\chi$)
 \begin{align} \label{eqa:msgl2_chi_rwa_p_dglsystem}
 \dot a&= -\bar{W} (c_1 b_1 - d_1) - \bar{T} (c_2 b_2 - d_2) -2(\bar{U} + \bar{V}) \notag \\
                    & \quad+ \Gamma(1-e^{i\chi})(\bar{B}^2 + \bar{D}^2)
                \,, \notag\\
 \dot b_1&=\bar{U} \cdot b_1- \bar{W} \cdot  b_1 \cdot  d_1 \,,\notag\\
 \dot b_2&= \bar{V} \cdot b_2 - \bar{T} \cdot b_2 \cdot d_2\,,\notag\\
  \dot c_1 &= \bar{U}\cdot c_1 - \bar{W} \cdot c_1 \cdot d_1 \,,\notag\\
 \dot c_2&= \bar{V} \cdot c_2  -\bar{T} \cdot c_2 \cdot d_2\,,\notag\\
 \dot d_1&= 2 \bar{U} \cdot d_1 - \bar{W} \cdot d_1^2 \notag \\
                        &\quad + \Gamma(1- e^{i\chi}) (\bar{A}^2 + \bar{B}^2)\,,\notag\\
  \dot  d_2&= 2 \bar{V} \cdot d_2 - \bar{T} \cdot d_2^2 \notag\\
                        &\quad+ \Gamma(1- e^{i\chi})(\bar{G}^2 + \bar{D}^2)\,.
 \end{align}
 The solution of this system in the long time limit is
 \begin{align}
\label{eqa:msgl2_chi_rwa_p_dglsys_lsg_t_gross}
d_1(\chi) &= \frac{U+\sqrt{U^2-\left(A^2+B^2\right) \left(-1+e^{i \chi }\right) W \Gamma }}{W} ;\notag \\
d_2(\chi) &=\frac{V+\sqrt{V^2-\left(-1+e^{i \chi }\right) \left(D^2+G^2\right) T \Gamma }}{T} ;\notag\\
b_1(\chi) &= b_2(\chi) = c_1(\chi) = c_2(\chi) = 0 \notag ;\\
a(t \to \infty ,\chi) &= \frac{1}{2} \biggl{(}-A^2+B^2+D^2-G^2 \notag\\
                & \quad +\sqrt{A^4+B^4-2 A^2 B^2 \left(-1+2 e^{2 i \chi }\right)} \notag\\
                &\quad +\sqrt{D^4-2 D^2 \left(-1+2 e^{2 i \chi }\right) G^2+G^4}\biggr{)} \Gamma t \,.
\end{align}
First, we find a steady state solutions of the last six equations, where time derivatives of $b_i,c_i,d_i$ $i \in {1,2}$ are zero. Then we insert them into the differential equation for $a(\chi,t)$, Eq. \eqref{eqa:msgl2_chi_rwa_p_dglsystem}, integrate, drop terms that are not dominant in case of $t \to \infty$ and obtain the equation for $a(t \to \infty, \chi)$.

\section{System: Average in  $P$-representation}
\label{cha:avarage_and_p}

In order to obtain $\avg{a_1^\dag a_1}$, we have to use the relation between the old and diagonal basis, because the P-representation has been evaluated in the diagonal basis.
\begin{align}
\avg{a_1^\dag a_1} &= \biggl{<} (B d_1 + A d_1^\dag + D d_2 + G d_2^\dag) \times \notag\\
                       & \quad \quad (A d_1 + B d_1^\dag + G d_2 + D d_2^\dag) \biggr{>}\,.
\intertext{We have computed the $P$-representation in the interaction picture using the RWA, within which we obtain} \avg{a_1^\dag a_1} &= \text{Tr} \int \bigl{(}\left(A^2 + B^2\right) d_1^\dag d_1 + \left(G^2 + D^2\right)d_2^\dag d_2  \notag \\
        & \quad\quad + B^2 + D^2\bigr{)}\cdot P \notag\\
        & \quad \quad \quad  \,\ketbra{\gamma_1}{\gamma_1} \otimes \ketbra{\gamma_2}{\gamma_2} \, d\,\gamma_1^2 d\,\gamma_2^2    \notag \\
  &= B^2 + D^2 + \int d\,\gamma_1^2 d\,\gamma_2^2 \Biggl{(}\left(A^2 + B^2\right) \gamma_1^* \gamma_1 \cdot P  \notag \\
  & \quad  + \left(G^2 + D^2\right)\gamma_2^* \gamma_2 \cdot P\Biggr{)} \,.\\
\intertext{The $P$-representation is normalized. With our ansatz for $P$ Eq. \eqref{eq1:msgl_rwa_p_dgl_loesung_ansatz} we can represent the terms like $\gamma_i \gamma_i^* P$ as  derivation with respect to $d_i$ such that $\gamma_i \gamma_i^* P = \partial_{d_i} P$  and do the integration $\int P d\gamma_1^2 d\gamma_2^2$ using symbolic numerical algebra to obtain.}
\avg{a_1^\dag a_1}(t) &= B^2 + D^2 + \left(A^2 + B^2\right) \frac{b_1(t) c_1(t) + d_1(t)}{d_1(t)^2} \notag \\
                   &\quad  + \left(G^2 + D^2\right) \frac{b_2(t) c_2(t) + d_2(t)}{d_2(t)^2}\,.
\end{align}

In a similar way,
\begin{align}\label{eq2:msgl_rwa_p_erw_allg_a2_mode}
    \avg{a_2^\dag a_2}(t) &= B_2^2 + D_2^2 + \left(A_2^2 + B_2^2\right)\frac{b_1(t) c_1(t) + d_1(t)}{d_1(t)^2} \notag \\
    & \quad + \left(G_2^2 + D_2^2\right)\frac{b_2(t) c_2(t) + d_2(t)}{d_2(t)^2}\,.
\end{align}



%

\end{document}